\documentclass[12pt]{article}
\usepackage{graphicx}
\usepackage{amssymb}
\usepackage{amscd}
\usepackage{amsmath}
\usepackage{cite}
\usepackage{xcolor}
\begin{document}
\begin{titlepage}

%\begin{center}
%{\hbox to\hsize{
%\hfill \bf hep-ph/??? }}
{\hbox to\hsize{\hfill September 2017 }}

\bigskip \vspace{3\baselineskip}

\begin{center}
{\bf \large 
Low temperature electroweak phase transition in the \\ Standard Model with hidden scale invariance}

\bigskip

\bigskip

{\bf Suntharan Arunasalam, Archil Kobakhidze, Cyril Lagger, \\ Shelley Liang and Albert Zhou \\ }

\smallskip

{ \small \it
ARC Centre of Excellence for Particle Physics at the Terascale, \\
School of Physics, The University of Sydney, NSW 2006, Australia 
\\}

\bigskip
 
\bigskip

\bigskip

{\large \bf Abstract}

\end{center}
\noindent 
We discuss a cosmological phase transition within the Standard Model which incorporates spontaneously broken scale invariance as a low-energy theory. In addition to the Standard Model fields, the minimal model involves a light dilaton, which acquires a large vacuum expectation value (VEV) through the mechanism of dimensional transmutation. Under the assumption of the cancellation of the vacuum energy, the dilaton develops a very small mass at 2-loop order. As a result, a flat direction is present in the classical dilaton-Higgs potential at zero temperature while the quantum potential admits two (almost) degenerate local minima with unbroken and broken eletroweak symmetry. We found that the cosmological electroweak phase transition in this model can only be triggered by a QCD chiral symmetry breaking phase transition at low temperatures, $T\lesssim 132$ MeV. Furthermore, unlike the standard case, the universe settles into the chiral symmetry breaking vacuum via a first-order phase transition which gives rise to a stochastic gravitational background with a peak frequency $\sim 10^{-8}$ Hz as well as triggers the production of approximately solar mass primordial black holes. The observation of these signatures of cosmological phase transitions together with the detection of a light dilaton would provide a strong hint of the fundamental role of scale invariance in particle physics. 
 \end{titlepage}

\section{Introduction}

Scale invariance provides an attractive framework for addressing the problem of the origin of mass and hierarchies of mass scales. In this framework, quantum fluctuations result in an overall mass scale via the mechanism of dimensional transmutation \cite{Coleman:1973jx}, while dimensionless couplings are responsible for generating mass hierarchies. The dimensionless couplings in the low-energy sector of the theory are only logarithmically sensitive to the high-energy sector and can be naturally small in the technical sense \cite{Wetterich:1983bi, Bardeen:1995kv, Kobakhidze:2014afa}.  If high-energy and low-energy sectors interact via feeble interactions, the breaking of scale invariance in the higher energy sector would proliferate in the low-energy sector resulting in a stable mass hierarchy between the two [for an incomplete list of recent works, see \cite{Foot:2007as, Meissner:2006zh}]. The above scenario is signified by the fact that scale (conformal) invariance is indeed an essential symmetry in string theory that is believed to provide a consistent ultraviolet completion of all fundamental interactions including gravity.

Recently, two of us have proposed a minimal extention of the Standard Model which incorporates spontaneously broken scale invariance as a low energy effective theory \cite{Kobakhidze:2017eml}. In this approach, non-linearly realised scale invariance is introduced  by promoting physical mass parameters (including the ultraviolet cut-off $\Lambda$) to a dynamical dilaton field. The dilaton field develops a large vacuum expectation value (VEV) via the quantum mechanical mechanism of dimensional transmutation. The dilaton-Higgs interactions then trigger the electroweak symmetry breaking and generate a stable hierarchy between the Higgs and dilaton VEVs. As a result of the spontaneous breaking of anomalous scale symmetry, the dilaton develops a mass at two loop level, which can be as small as $\sim 10^{-8}$ eV (for a dilaton VEV of the order the Planck scale, $\sim M_P\sim 10^{19}$ GeV). In addition, the Higgs-dilaton potential displays  a nearly flat direction. 

The formalism of hidden scale invariance is rather generic and can be applied to other effective field theory models, with essentially the same predictions regarding the light dilaton and the Higgs-dilaton potential \cite{Kobakhidze:2017mcz}. Due to these generic features it is interesting to investigate the cosmological phase transition in effective theories with hidden scale invariance. This is the purpose of the present paper. 

Witten has pointed out  a long time ago \cite{Witten:1980ez} that in the Standard Model with Coleman-Weinberg radiative electroweak symmetry breaking, the cosmological electroweak phase transition is strongly first-order. The electroweak phase transition is aided by the QCD quark-antiquark condensate and hence occurs at low temperatures, namely around the temperature of the QCD chiral phase transition. See also the follow up work which also introduces the dilaton field \cite{Buchmuller:1990ds}. Although these models are no longer phenomenologically viable, one may consider their extensions which exhibit the same features for some range of parameters \cite{Iso:2017uuu}. We will argue below, that within the framework of hidden scale invariance, the electroweak phase transition is necessarily triggered by QCD chiral phase transition and is completed at a low temperature $\sim 130$ MeV. Unlike the previously discussed models, we find that the Higgs field transitions to the electroweak vacuum via a second-order phase transition, while the chiral phase transition becomes first-order. The later phase transition leads to the generation of stochastic gravitational waves in the $\sim 10^{-8}$ Hz frequency range, which are potentially observable using pulsar timing technique, e.g. at the Square Kilometre Array (SKA) observatory \cite{Huynh:2013aea}. In addition, production of primordial solar mass black holes are expected during that phase transition. 

The paper is organised as follows. In the next section we describe the minimal Standard Model with hidden scale invariance. Calculation of the thermal effective potential and a subsequent analysis of the cosmological phase transition is given is section 3. The last section 4 is reserved for conclusions.

\section{The Standard Model with hidden scale invariance}

Let us consider the Standard Model as an effective low energy theory valid up to an energy scale, $\Lambda$, as introduced in \cite{Kobakhidze:2017eml}. In the Wilsonian approach, the ultraviolet cut-off $\Lambda$ is a physical parameter that encapsulates physics (e.g. massive fields) which we are agnostic of. The Higgs potential defined at this ultraviolet scale reads:
\begin{equation}
V(\Phi^{\dagger}\Phi)=V_0(\Lambda)+\lambda(\Lambda)\left[\Phi^{\dagger}\Phi - v_{ew}^2(\Lambda)\right]^2
+...,
\label{1}
\end{equation}       
where $\Phi$ is the electroweak doublet Higgs field, $V_0$ is a field-independent constant (bare cosmological constant parameter) and the ellipsis denote all possible dimension $> 4$ (irrelevant), gauge invariant operators, $\left(\Phi^{\dagger}\Phi\right)^n$, $n=3,4...$. The other bare parameters include the dimensionless couplings $\lambda(\Lambda)$ and a mass dimension parameter $v_{ew}(\Lambda)$ namely the bare Higgs expectation value. In principle, this potential has an infinite number of nonrenormalisable operators and $\Lambda$-dependent parameters must fully encode the physics beyond the Standard Model. In practice, however, we usually deal with a truncated theory, which is valid in the low-energy domain only. 

We assume now that a fundamental theory maintains spontaneously broken scale invariance, such that all mass parameters  have a common origin. To make this symmetry manifest in our effective theory, we promote all mass parameters to a dynamical field $\chi$, the dilaton, as follows:
\begin{equation}
\Lambda \to \Lambda \frac{\chi}{f_{\chi}}\equiv \alpha \chi,~~v_{ew}^2(\Lambda) \to \frac{v_{ew}^2(\alpha\chi)}{f_{\chi}^2}\chi^2\equiv \frac{\xi (\alpha\chi)}{2}\chi^2,~~V_0(\Lambda)\to \frac{V_0(\alpha\chi)}{f_{\chi}^4}\chi^4\equiv \frac{\rho(\alpha\chi)}{4}\chi^4~,  
\label{2}
\end{equation}  
where $f_{\chi}$ is the dilaton decay constant. Then, Eq. (\ref{1}) turns into the Higgs-dilaton potential, 
 \begin{equation}
V(\Phi^{\dagger}\Phi, \chi)=\lambda(\alpha\chi)\left[\Phi^{\dagger}\Phi  -\frac{\xi(\alpha\chi)}{2}\chi^2  \right]^2 +\frac{\rho(\alpha\chi)}{4}\chi^4~.  
\label{3}
\end{equation}  
This potential is manifestly scale invariant up to the quantum scale anomaly, which is engraved in the $\chi$-dependence of dimensionless couplings\footnote{In this we differ substantially from the so-called quantum scale-invariant SM \cite{Shaposhnikov:2008xi}. In their approach, the SM is extrapolated to an arbitrary high energy scale and regularized by invoking dilaton-dependent renormalization scale, $\mu=\mu(\chi)$}. Indeed, the Taylor expansion around an arbitrary fixed scale $\mu$ reads:
 \begin{equation}
\lambda^{(i)}(\alpha\chi)=\lambda^{(i)}(\mu)+\beta_{\lambda^{(i)}}(\mu)\ln\left(\alpha\chi/\mu\right)+\beta'_{\lambda^{(i)}}(\mu)\ln^2\left(\alpha\chi/\mu\right)+...,
\label{4}
\end{equation}  
where $\lambda^{(i)} \equiv (\lambda, \xi, \rho)$ and
 \begin{equation}
\beta_{\lambda^{(i)}}(\mu)=\left. \frac{\partial \lambda^{(i)}}{\partial \ln\chi}\right |_{\alpha\chi=\mu}~,
\label{5}
\end{equation}  
is the renormalisation group (RG) $\beta$-functions for the respective coupling $\lambda^{(i)}$ defined at a scale $\mu$, while $\beta'_{\lambda^{(i)}}(\mu)=\left. \frac{\partial^2 \lambda^{(i)}}{\partial (\ln\chi )^2}\right |_{\alpha\chi=\mu}$, etc. For convenience, we fix the renormalisation scale at the cut-off scale $\Lambda$, which is defined through the  dilaton VEV as $\langle \chi \rangle \equiv v_{\chi}$, i.e. $\mu=\Lambda=\alpha v_{\chi}$. Note that while the lowest order contribution in $\beta$-functions is one-loop, i.e. $\sim {\cal O}(\hbar)$, $n$-th derivative of $\beta$ is $nth$ order  in the perturbative loop expansion, $\sim {\cal O}(\hbar^n)$.  

The extremum condition $\left. \frac{dV}{d\chi}\right |_{\Phi=\langle\Phi\rangle, \chi=\langle\chi\rangle}=0$ together with the phenomenological constraint on vacuum energy $V(v_{ew}, v_{\chi})=0$, lead to the following relations:
\begin{eqnarray}
\rho(\Lambda)=0~,~~ \beta_{\rho}(\Lambda)=0~.
\label{6}
\end{eqnarray}  
One of the above  relations can be used to define the dilaton VEV (dimensional transmutation) and another represents the tuning of the cosmological constant.  The second extremum condition  $\left. \frac{dV}{d\Phi}\right |_{\Phi=\langle\Phi\rangle, \chi=\langle\chi\rangle}=0$ simply sets the hierarchy of VEVs:
 \begin{eqnarray}
\xi(\Lambda) = \frac{v^2_{ew}}{v^2_{\chi}}~.
\label{7}
\end{eqnarray} 
In the classical limit when all the quantum corrections are zero, i.e., $\beta_{\lambda^{(i)}}=\beta'_{\lambda^{(i)}}=...=0$, the above vacuum configuration represents a flat direction of the Higgs-dilaton potential (\ref{3}). The existence of this flat direction is, of course, the direct consequence of the assumed classical scale invariance. In this approximation, the dilaton is the massless Goldstone boson of spontaneously broken scale invariance. The flat direction is lifted by quantum effects and, as we will see below, by thermal effects in the early universe. Note, however, that the dilaton develops a (running) mass in our scenario at two-loop level \cite{Kobakhidze:2017eml} (see also \cite{Foot:2010et}),  
\begin{eqnarray}
m_{\chi}^2\simeq\frac{\beta'_{\rho}(\Lambda)}{4\xi(\Lambda)}v^2_{ew}~,
\label{8}
\end{eqnarray} 
 while the tree-level Higgs mass is given to a high accuracy by the standard formula: $m_h^2\simeq2\lambda (\Lambda) v^2_{ew}$. Note that $\beta'_{\rho}\propto \xi^2$ and hence the dilaton is a very light particle, $m_{\chi}/m_h\sim \sqrt{\xi}$. 
 
\begin{figure}[t!]
    \centering
    \includegraphics[width=0.7\textwidth]{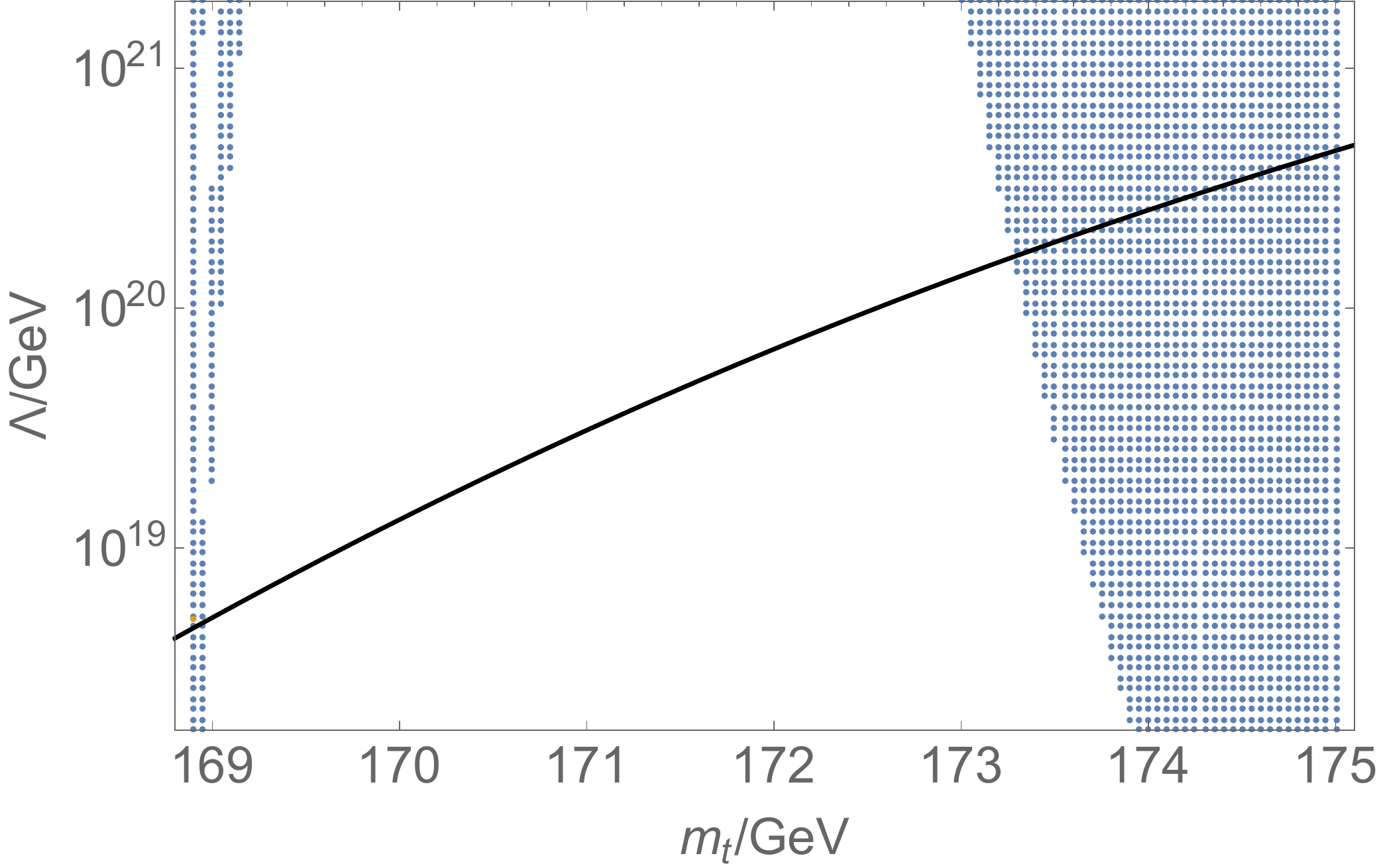}%
    \vspace{-10pt}
    \caption{\small Plot of the allowed range of parameters (shaded region) with $m_{\chi}^2(v_{ew})>0$, i.e., the electroweak vacuum being a minimum. The solid line displays the cut-off scale $\Lambda$ as function of the top-quark mass $m_t$ for which the conditions in Eq. (\ref{6}) are satisfied.}
    \label{fig1_mass}
\end{figure}

To verify whether the above scalar field configurations correspond to a local minimum of the potential one must evaluate the running masses down to low energy scales. The relations in Eq. (\ref{6}) provide non-trivial constraints here. In Figure \ref{fig1_mass}, we have presented our analysis based on solutions of the relevant (one-loop) RG equations (see the appendix section in Ref. \cite{Kobakhidze:2017eml}). The shaded region in the $\Lambda - m_t$ plane corresponds to a positive dilaton mass squared (minimum of the potential) and the solid curve shows the cut-off scale $\Lambda$ as a function of the top-quark mass $m_t$ for which the conditions in Eq. (\ref{6}) are satisfied. Hence, within the given approximation, we find that the model is phenomenologically viable for $m_t\lesssim 169$ GeV and $m_t\gtrsim 173$ GeV with the cut-off scale accordingly predicted to be $\Lambda \lesssim 10^{19}$ GeV and $\gtrsim 10^{21}$ GeV respectively. We note that the upper values are within the allowed experimental range for the top quark mass, $m_t=173.34\pm0.27(\text{stat})\pm0.71(\text{syst})\ \text{GeV}$ \cite{ATLAS:2014wva}. Assuming $v_{\chi}\approx \Lambda~( \alpha\approx 1)$, the dilaton mass for the Planck scale cut-off is predicted to be $m_{\chi}\approx 10^{-8}$ eV. This prediction for the ultraviolet scale $\Lambda$, however, should be taken as indicative only. Indeed, besides high-loop corrections, the actual matching conditions (threshold effects) between low energy couplings and couplings in the ultraviolet completion of the Standard Model may affect the above predictions significantly (see, e.g., examples in Ref. \cite{Kobakhidze:2013pya}). However, these details of the evaluation of coupling constants at high energy scales are not essential for the purpose of the present study of electroweak phase transition. In what follows we assume $\Lambda=M_P\approx 10^{19}$ GeV in our numerical calculations. 
%%%%%%%%%%%%% FIGURE 1 %%%%%%%%%%%%%%%%%%%%%%%%%%%% 
%%%%%%%%%%%%%%%%%%%%%%%%%%%%%%%%%%%%%%%%%%%%%%%

Another important observation is that, the potential energy densities evaluated at the origin and at the electroweak minimum are equal, $V(0,0)\simeq V(v_{ew}, v_{\chi})$. This is readily seen for the potential evaluated at the cut-off scale $\Lambda$, see Eqs. (\ref{3}) and (\ref{6}, \ref{7}). Then, since the vacuum energy density does not depend on the renormalisation scale \cite{Foot:2007wn}, the trivial and electroweak vacuum states must be degenerate at any given low energy scale. This has an important ramification for the cosmological phase transition - the critical temperature of the electroweak phase transition, $T_c$, defined as the temperature where the two minima are degenerated, is $T_c=0$. Hence, the premature conclusion is disastrous for our model: no electroweak phase transition is possible. However, as it will be shown in the next section, as the universe cools down QCD chiral symmetry breaking happens such that the quark-antiquark condensate triggers the electroweak symmetry breaking and the Higgs field relaxes in its electroweak symmetry breaking vacuum configuration.

\section{The electroweak phase transition}

In this section, we discuss the electroweak phase transition within the scale-invariant Standard Model described above. To this end, we adopt the standard method (for a review see, e.g., Ref. \cite{Quiros:1999jp}), treating the potential in Eq. (\ref{3}) evaluated at the cut-off scale $\Lambda$ as a tree-level potential, as  is common for Wilsonian effective theories. A notable difference from the standard formalism is the presence of the dilaton-dependent dynamical cut-off in our case. The standard quadratic divergent term $\propto \Lambda^2$, which are renormalised away within the standard calculations,  and quartic (field-independent) divergent terms $\propto \Lambda^4$, which are typically ignored altogether, become now $\propto \chi^2$ and $\propto \chi^4$. Purely quantum (temperature-independent) corrections of this sort can be absorbed in the redefinition of the tree-level couplings in Eq. (\ref{3}). The  temperature-independent logarithmic terms $\propto \ln \chi^2$ in our calculations do exactly reproduce the standard zero temperature Coleman-Weinberg quantum corrections. As discussed in the previous section, quantum corrections break explicitly scale invariance and give rise to the dilaton mass at two-loop level. Therefore,  in the early universe they are subdominant compared to the thermal corrections (especially along the classical flat direction), which also break scale invariance explicitly.  Thus, we can safely ignore the quantum corrections in what follows. The leading high temperature expansion of the effective finite temperature Higgs-dilaton potential then reads: 
\begin{eqnarray}
V_T(h, \chi)&=&\frac{\lambda(\Lambda)}{4}\left[h^2-\frac{v_{ew}^2}{v_{\chi}^2}\chi^2\right]^2 \nonumber\\
&+&c(h)\pi^2T^4-\frac{\lambda(\Lambda)}{24}\frac{v_{ew}^2}{v_{\chi}^2}\chi^2T^2+
\frac{1}{48}\left[6\lambda (\Lambda)+6y_t^2(\Lambda)+
\frac{9}{2}g^2(\Lambda)+\frac{3}{2}g'^2(\Lambda)\right]h^2T^2~
\label{9}
\end{eqnarray} 
 where $h$ is the neutral, CP-even component of the Higgs doublet, $\Phi=\left(0, h/\sqrt{2}\right)^{\rm T}$ and $c(h)$ is a number of relativistic degrees of freedom, which are in thermal equilibrium at $T$. The parameter $c(h)$ has implicit $h$ dependence, through the relation $m_i(h)<T$, where $m_i(h)$ are $h$-dependent masses for Standard Model fields. Only dominant thermal fluctuations of heaviest Standard Model fields ($i=W^{\pm}, Z, h, t$) are taken into account and the relations (\ref{6}) and (\ref{7}) are employed when deriving Eq. (\ref{9})
 
 To proceed further, we first eliminate the dilaton field by solving its equation of motion, $\partial V_T/\partial \chi =0$, which implies at leading order:
 \begin{eqnarray}
\chi^2 \approx \frac{v_{\chi}^2}{v_{ew}^2}\left(h^2+\frac{T^2}{12}\right)~. 
\label{10}
\end{eqnarray}
Note that if we set the temperature to zero, the above equation displays the flat direction of the zero temperature classical potential. Hence, the $T^2$ term is the leading contribution from thermal fluctuations that breaks scale invariance explicitly. Plugging in (\ref{10}) back into (\ref{9}), we obtain the finite temperature potential in terms of the Higgs field only:   
 \begin{eqnarray}
V_T(h, \chi(h))&=&\left[c(h)\pi^2 -\frac{\lambda(\Lambda)}{576}\frac{v_{ew}^2}{v_{\chi}^2}(2+v_{ew}^2/v_{\chi}^2)\right]T^4 \nonumber \\ 
&+&
\frac{1}{48}\left[4\lambda (\Lambda)+6y_t^2(\Lambda)+
\frac{9}{2}g^2(\Lambda)+\frac{3}{2}g'^2(\Lambda)\right]h^2T^2~
\label{11}
\end{eqnarray} 
As expected in this approximation, the temperature independent terms vanish due to the flatness of the classical potential. We verified numerically that $4\lambda (\Lambda)+6y_t^2(\Lambda)+\frac{9}{2}g^2(\Lambda)+\frac{3}{2}g'^2(\Lambda)>0$ and therefore the curvature of the effective potential (\ref{11}) at the origin $h=0$ is positive. Hence, $h=0$ is a minimum of the effective potential and is separated from another local minimum, which corresponds to the electroweak symmetry breaking configuration of the Higgs field by the temperature dependent barrier. Furthermore, this barrier persists down to $T=0$ due to the two (and higher) loop quantum corrections with, as it has been discussed early, two vacuum states being degenerated in energy. We stress again that this generic prediction of the model is largely independent on its ultraviolet completion and would imply that the universe is stuck in the trivial symmetric minimum. 

However, the above picture is actually altered as the universe cools down to temperatures where QCD interactions become strong and various composite states start to form. As the universe remains in the symmetric phase $h=0$, all quarks (and other Standard Model particles) are massless at that epoch. Hence, the $SU(6)_L\times SU(6)_R$ chiral symmetry in the quark sector must be exact and it gets spontaneously broken once QCD quark-antiquark condensate forms. Part of the $SU(6)_L\times SU(6)_R$ chiral symmetry is actually gauged and represents $SU(2)\times U(1)$ electroweak symmetry. Therefore, the quark-antiquark condensate also breaks the electroweak symmetry and results in generation of small masses for the  $W^{\pm}$ and $Z$ gauge bosons. The finite temperature quark-antiquark condensate, $\langle \bar q q\rangle_T$ has been computed within the chiral perturbation theory with $N$ massless quarks in \cite{Gasser:1986vb}:
\begin{eqnarray}
\langle \bar q q\rangle_T=\langle \bar q q\rangle\left[1-(N^2-1)\frac{T^2}{12Nf^2_{\pi}}-\frac{1}{2}
(N^2-1)\left(\frac{T^2}{12Nf^2_{\pi}}\right)^2+{\cal O}\left((T^2/12Nf^2_{\pi})^3\right)\right]~,
\label{12}
\end{eqnarray}
 where $\langle \bar q q\rangle \approx -(250$ MeV)$^3$ is the zero temperature condensate and $f_{\pi}\approx 93$ MeV is the pion decay constant. From Eq. (\ref{12}) we can infer that for $N=6$ the critical temperature of the chiral symmetry breaking phase transition, defined by $\langle \bar q q\rangle_{T_c}=0$, is equal to $T_c\approx 132$ MeV. The condensate (\ref{12}) would generate a linear term in the effective potential through the quark-Higgs Yukawa interactions: $y_q\langle \bar q q\rangle_{T}h/\sqrt{2}$, where $y_q$ is the Yukawa coupling of $q-$type quark. It should be stressed that while all terms in the effective potential Eq. (\ref{11}) diminish as $T$ decreases, the magnitude of the linear term increases. 
The extremum condition is modified as: 
\begin{eqnarray}
y_q\langle \bar q q\rangle_{T}/\sqrt{2}+\frac{\partial V_T}{\partial h}=0~, 
\label{13}
\end{eqnarray}
 and it is clear that the local minimum shifts from the origin $h=0$ to non-zero values of $h$.
 
\begin{figure}[t!]
    \centering
    \includegraphics[width=0.7\textwidth]{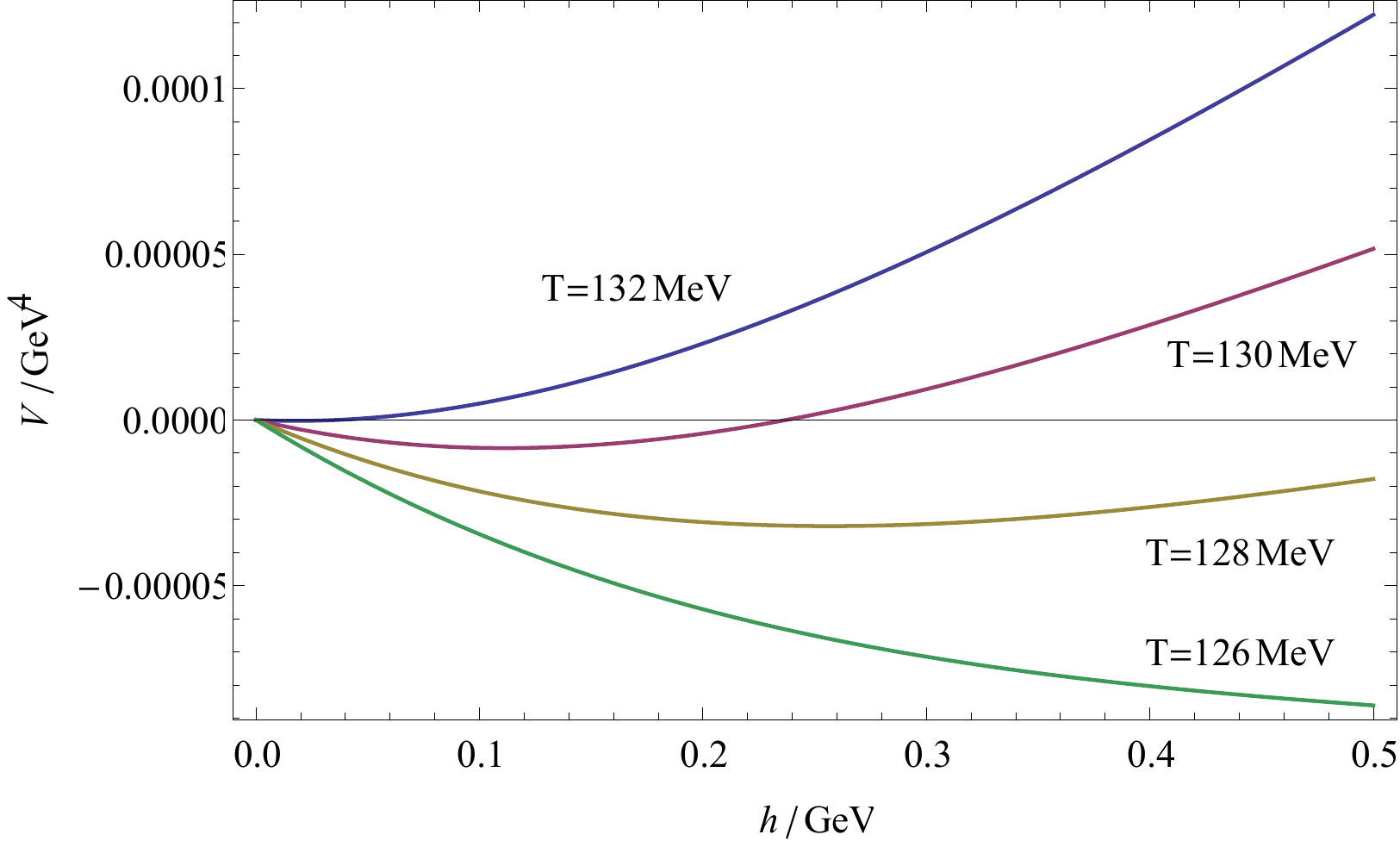}%
   % \vspace{-64pt}
   \caption{\small $V_T(h)-V_T(0)$ for different temperatures below the chiral phase transition.}
    \label{chiral}
\end{figure}

We have analysed the evolution of the local minimum numerically by employing the full finite temperature effective potential (see Appendix \ref{app}). Just below the critical point of the chiral phase transition at $T_c$, the QCD condensate term is small and a non-zero minimum, $h_0$ does emerge near $h=0$ (see Figure \ref{chiral}). This minimum is separated by a potential barrier from another local minimum that later evolves into the eletroweak vacuum. This minimum exists for $127~{\rm GeV}\leq T \leq 132~{\rm GeV}$. In this range of temperatures, the top quark remains relativistic with $\frac{m_t(h_0)}{T}\lesssim 1 $.  Below this range of temperatures, the contribution from $|\langle \bar q q\rangle_T|$  becomes large enough such that the local minimum near the origin no longer occurs, indeed the first term in Eq. (\ref{13}) becomes larger than the second term. Subsequently, the Higgs field quickly runs down classically the slope from near the origin towards the true electroweak breaking vacuum.  
%%%%%%%%%%%%% FIGURE 2 %%%%%%%%%%%%%%%%%%%%%%%%%%%% 
%%%%%%%%%%%%%%%%%%%%%%%%%%%%%%%%%%%%%%%%%%%%%%%

In contrast with the previous studies \cite{Iso:2017uuu}, the change of the Higgs field configuration during this kind of phase transition is smooth and homogeneous, and does not proceed through $h-$bubble nucleation. However, since the QCD phase transition precedes the electroweak one, all the six flavours of quarks are massless during that process. There are theoretical arguments \cite{Pisarski:1983ms}, which are supported by numerical calculations \cite{Gausterer:1988fv}, which suggest that the QCD phase transition with $N\geq 3$ massless quarks is first-order. If true, this may have interesting cosmological consequences within our model as briefly outlined below.

First-order cosmological phase transitions are known to be a source of gravitational waves, which are generated through the dynamics of vacuum bubbles. We can readily estimate the characteristics of these gravitational waves during the first-order QCD phase transition discussed above. First, note that during the whole phase transition the universe is dominated by radiation since the energy difference between the two vacua is due to the difference in thermal energy (the $\propto c(h)T^4$ term in Eq. (\ref{11})), which vanishes as $T\to 0$. Therefore the peak frequency (observed today) of the waves produced at $T_c\approx 132$ MeV can be estimated as: $f\sim H(T_c)\frac{T_0}{T_c}\sim \frac{T_cT_0}{M_P}\sim 10^{-8}$ Hz ($T_0\approx 10^{-4}$ eV). The stochastic gravitational waves background with such frequencies can potentially be detected by means of pulsar timing arrays \cite{Caprini:2010xv}\footnote{Similar prediction has been made within a model based on a prolonged electroweak phase transition in Ref. \cite{Kobakhidze:2017mru}}, e.g. with the SKA telescope \cite{Huynh:2013aea}. 

Another interesting phenomenon associated with a first-order QCD phase transition is the production of primordial black holes \cite{Crawford:1982yz}. The mass of a horizon size black hole can be estimated to be of the order of solar mass, $M_{bh}\sim \frac{1}{H(T_c)G_N}\sim \frac{M_P^3}{T_c^2}\sim 10^{65}~{\rm eV}\sim M_{\odot}$. These are large enough black holes to survive the Hawking evaporation until the present epoch and thus can contribute to the total dark matter density.  Finally, we also mention that our model may provide a natural framework for the so-called cold baryogenesis \cite{Servant:2014bla}. The observation of such signatures of a cosmological phase transition together with the detection of a light dilaton would provide a strong hint of the fundamental role of scale invariance in particle physics. 
     
\section{Conclusion}

In this paper we have studied a cosmological electroweak phase transition within the minimal Standard Model with hidden scale invariance. The model predicts a light dilaton which very feebly couples to the Standard Model fields. The Higgs-dilaton potential exhibits two degenerate minima at zero temperature, therefore the electroweak phase transition can only be triggered by the QCD chiral symmetry phase transition at $T\lesssim 132$ MeV. We found that the Higgs field configuration changes smoothly during this transition, while the chiral symmetry breaking is likely to be first-order. Consequently, gravitational waves with peak frequency $\sim 10^{-8}$ Hz and stable primordial black holes of mass $\sim M_{\odot}$ are predicted to be produced during the phase transition. We plan more detailed investigation of these phenomena in the forthcoming publication \cite{new}. 

\paragraph{Acknowledgement.} The work was supported in part by the Australian Research Council. CL would like to thank Mikhail Shaposhnikov for his hospitality at the Laboratory of Particle Physics and Cosmology at EPF Lausanne, where a part of this research has been conducted. AK completed this work while visiting the workshop at the Galileo Galilei Institute "Collider Physics and the Cosmos".

\appendix
\label{app}
\section{Calculation of the finite temperature effective potential}
The contribution of a scalar field with field-dependent mass, $m(h)$, to the thermal effective potential with a 4-dimensional cut-off, \(\omega_n^2+\vec{p}^2\leq\Lambda^2\) is given by:
\begin{equation}
	V_T(h) = \frac{1}{2\beta}\int_{|\vec{p}|\leq\Lambda} \frac{d^3p}{(2\pi)^3}\sum_{n=-N}^{N}\log\left(1+\frac{m^2(h)}{\omega_n^2+\vec{p}^2}\right),
	\end{equation}
	where \(N=\left\lfloor\frac{\beta}{2\pi}\sqrt{\Lambda^2-\vec{p}^2}\right\rfloor\), \(\beta=\frac{1}{T}\) and   \(\omega_n = \frac{2n\pi}{\beta}\). 
	Define 
	\begin{equation}
		v(y) = \sum_{n=-N}^{N} \log(n^2+y^2)
	\end{equation}
	and \(\omega^2=p^2+m^2(h)\). 
	The one-loop contribution then becomes
	\begin{equation}
			V_T(h)  = \frac{1}{2\beta}\int_{|\vec{p}|\leq\Lambda} \frac{d^3p}{(2\pi)^3}v\left(\frac{\beta\omega}{2\pi}\right) - \frac{1}{2\beta}\int_{|\vec{p}|\leq\Lambda} \frac{d^3p}{(2\pi)^3}v\left(\frac{\beta p}{2\pi}\right). \label{app2}
	\end{equation}
	The difference of these integrals can thus be found by evaluating the first integral and discarding any $m$-indepenedent terms. By using the fact that \(\log(n^2+y^2)=\log(n+iy)+\log(n-iy)\), it can be shown using the properties of Gamma functions as well as the Euler reflection formula that:
	\begin{align}
		v(y) = 2\log\sinh \pi y + 4\Re\log\Gamma(1+N+iy)+ \text{\(m\)-independent terms}\label{app1}. 
		\end{align}
		The first term of Equation \ref{app1}, when integrated, yields:
		\begin{align}\frac{1}{2\beta}\int_{|\vec{p}|\leq\Lambda} \frac{d^3p}{(2\pi)^3}\left(2\log\sinh\left(\frac{\beta\omega}{2}\right) - 2\log\sinh\left(\frac{\beta p}{2}\right)\right)\nonumber\\
		=\frac{1}{2\pi^2}\left[\frac{1}{8}m^2\Lambda^2 + \left(\frac{1}{4}-\log 2-\log\frac{\Lambda}{m}\right)\frac{m^4}{16} 
		+\frac{1}{\beta^4} J_B(m^2\beta^2)\right]\label{app4}
		\end{align}
		where terms with negative powers of $\Lambda$ are ignored and 
		\[J_B(m^2\beta^2)=\int_0^\infty dx x^2\log\left[1-e^{-\sqrt{x^2+\beta^2m^2}}\right].\]
		(In the above we take the upper integral limit to infinity, as contributions for large \(x\) to the integral decay exponentially.) Using the Schwarz reflection principle and the Stirling formula, the second term of Equation \ref{app1} can be expanded as 
		\begin{equation}
		4\Re\log\Gamma(1+z) = 4\Re z\log z + 2\Re\log z+\frac{1}{3}\Re\frac{1}{z}+O\left(\frac{1}{z^3}\right)+\text{\(m\)-independent terms}.
		\end{equation} where $z=N+iy$. Integrating this, one finds that the corresponding contribution to Equation \ref{app2} is 
		\begin{align}
		\frac{2\log 2-1}{64\pi^2}m^4 - \frac{1}{32\pi^2}m^2\Lambda^2+  O\left(\frac{1}{\beta\Lambda}\right)+O\left(\frac{m}{\Lambda}\right).\label{app3}
\end{align}
		The integration process is rather arduous, as \(N\) must be split into a continuous function and a sawtooth function. One then takes advantage of the fact that the sawtooth function has a small period to extract relevant terms of positive powers of \(\Lambda\).
		Hence, adding equations \ref{app4} and \ref{app3} together, one obtains:
		 \begin{equation}
			V_T(h)  = \frac{1}{32\pi^2}m^2\Lambda^2-\frac{1}{128\pi^2}m^4-\frac{1}{32\pi^2}\log\frac{\Lambda}{m}m^4+ \frac{1}{2\pi^2}T^4J_B(m^2/T^2)\label{boson}
	\end{equation}
		Similarly, for a fermion field, one obtains the contribution:
	 \begin{equation}
			V_T(h)  =- \frac{1}{32\pi^2}m^2\Lambda^2+\frac{1}{128\pi^2}m^4+\frac{1}{32\pi^2}\log\frac{\Lambda}{m}m^4+ \frac{1}{2\pi^2}T^4J_F(m^2/T^2)\label{fermion}
	\end{equation}
	where 
	\[J_F(m^2\beta^2)=\int_0^\infty dx x^2\log\left[1+e^{-\sqrt{x^2+\beta^2m^2}}\right]\].
	Now, as $\Lambda$ is proportional to $\chi$ in a scale invariant model, the first two terms of Equations \ref{boson} and \ref{fermion} can be included into the tree level potential through the redefinition of $\xi$ and $\lambda$.  Hence, the full thermal effective potential is given by:
	\begin{align}
	V_T(h,\chi)=\frac{\lambda(\Lambda)}{4}\left[h^2-\frac{v_{ew}^2}{v_{\chi}^2}\chi^2\right]^2+\sum_{i}n_i(-1)^{2s_i+1}\left[\frac{m_i^4}{32\pi^2}\log\frac{\alpha \chi}{m_i}- \frac{1}{2\pi^2}T^4J_i(m_i^2/T^2)\right]
	\end{align}
	where $i$ runs over all relativistic particles, $n_i$ is the number of degrees of freedom of the corresponding particle, $s_i$ is the spin and $J_i(y)$ is $J_B(y)$ for bosons and  $J_F(y)$ for fermions.


\begin{thebibliography}{999}

  %\cite{Coleman:1973jx}
\bibitem{Coleman:1973jx} 
  S.~R.~Coleman and E.~J.~Weinberg,
  %``Radiative Corrections as the Origin of Spontaneous Symmetry Breaking,''
  Phys.\ Rev.\ D {\bf 7}, 1888 (1973).
  doi:10.1103/PhysRevD.7.1888
  %%CITATION = doi:10.1103/PhysRevD.7.1888;%%
  %3885 citations counted in INSPIRE as of 09 Jul 2017

%\cite{Wetterich:1983bi}
\bibitem{Wetterich:1983bi} 
  C.~Wetterich,
  %``Fine Tuning Problem And The Renormalization Group,''
  Phys.\ Lett.\ B {\bf 140}, 215 (1984); 
  %%CITATION = PHLTA,B140,215;%%
  %19 citations counted in INSPIRE as of 19 May 2013

%\cite{Bardeen:1995kv}
\bibitem{Bardeen:1995kv}
  W.~A.~Bardeen,
  %``On naturalness in the standard model,''
  FERMILAB-CONF-95-391-T.
  %%CITATION = FERMILAB-CONF-95-391-T;%%

%\cite{Kobakhidze:2014afa}
\bibitem{Kobakhidze:2014afa} 
  A.~Kobakhidze and K.~L.~McDonald,
  %``Comments on the Hierarchy Problem in Effective Theories,''
  JHEP {\bf 1407}, 155 (2014)
  %doi:10.1007/JHEP07(2014)155
  [arXiv:1404.5823 [hep-ph]].
  %%CITATION = doi:10.1007/JHEP07(2014)155;%%
  %10 citations counted in INSPIRE as of 31 Dec 2015

%\cite{Foot:2007as}
\bibitem{Foot:2007as} 
  R.~Foot, A.~Kobakhidze and R.~R.~Volkas,
  %``Electroweak Higgs as a pseudo-Goldstone boson of broken scale invariance,''
  Phys.\ Lett.\ B {\bf 655}, 156 (2007)
  doi:10.1016/j.physletb.2007.06.084
  [arXiv:0704.1165 [hep-ph]];
  %%CITATION = doi:10.1016/j.physletb.2007.06.084;%%
  %107 citations counted in INSPIRE as of 12 Nov 2016
  %\cite{Foot:2010av}
%\bibitem{Foot:2010av} 
 % R.~Foot, A.~Kobakhidze and R.~R.~Volkas,
  %``Stable mass hierarchies and dark matter from hidden sectors in the scale-invariant standard model,''
  Phys.\ Rev.\ D {\bf 82}, 035005 (2010)
  doi:10.1103/PhysRevD.82.035005
  [arXiv:1006.0131 [hep-ph]];
  %%CITATION = doi:10.1103/PhysRevD.82.035005;%%
  %49 citations counted in INSPIRE as of 12 Nov 2016
  %\cite{Foot:2007ay}
%\bibitem{Foot:2007ay} 
  R.~Foot, A.~Kobakhidze, K.~L.~McDonald and R.~R.~Volkas,
  %``Neutrino mass in radiatively-broken scale-invariant models,''
  Phys.\ Rev.\ D {\bf 76}, 075014 (2007)
  doi:10.1103/PhysRevD.76.075014
  [arXiv:0706.1829 [hep-ph]]; 
  %%CITATION = doi:10.1103/PhysRevD.76.075014;%%
  %81 citations counted in INSPIRE as of 12 Nov 2016
  %\cite{Foot:2007iy}
%\bibitem{Foot:2007iy} 
  %R.~Foot, A.~Kobakhidze, K.~L.~McDonald and R.~R.~Volkas,
  %``A Solution to the hierarchy problem from an almost decoupled hidden sector within a classically scale invariant theory,''
  Phys.\ Rev.\ D {\bf 77}, 035006 (2008)
  doi:10.1103/PhysRevD.77.035006
  [arXiv:0709.2750 [hep-ph]];
  %%CITATION = doi:10.1103/PhysRevD.77.035006;%%
  %149 citations counted in INSPIRE as of 12 Nov 2016
%\cite{Foot:2013hna}
%\bibitem{Foot:2013hna} 
 % R.~Foot, A.~Kobakhidze, K.~L.~McDonald and R.~R.~Volkas,
  %``Poincaré protection for a natural electroweak scale,''
  Phys.\ Rev.\ D {\bf 89}, no. 11, 115018 (2014)
  doi:10.1103/PhysRevD.89.115018
  [arXiv:1310.0223 [hep-ph]].
  %%CITATION = doi:10.1103/PhysRevD.89.115018;%%
  %41 citations counted in INSPIRE as of 31 Dec 2015
%\cite{ATLAS:2014wva}
\bibitem{ATLAS:2014wva}
  [ATLAS and CDF and CMS and D0 Collaborations],
  %``First combination of Tevatron and LHC measurements of the top-quark mass,''
  arXiv:1403.4427 [hep-ex].
  %%CITATION = ARXIV:1403.4427;%%
  %478 citations counted in INSPIRE as of 13 Sep 2017

%\cite{Meissner:2006zh}
\bibitem{Meissner:2006zh} 
  K.~A.~Meissner and H.~Nicolai,
  %``Conformal Symmetry and the Standard Model,''
  Phys.\ Lett.\ B {\bf 648}, 312 (2007)
  doi:10.1016/j.physletb.2007.03.023
  [hep-th/0612165];
  %%CITATION = doi:10.1016/j.physletb.2007.03.023;%%
  %217 citations counted in INSPIRE as of 12 Nov 2016
%\cite{Iso:2009ss}
%\bibitem{Iso:2009ss} 
  S.~Iso, N.~Okada and Y.~Orikasa,
  %``Classically conformal $B^-$ L extended Standard Model,''
  Phys.\ Lett.\ B {\bf 676}, 81 (2009)
  doi:10.1016/j.physletb.2009.04.046
  [arXiv:0902.4050 [hep-ph]];
  %%CITATION = doi:10.1016/j.physletb.2009.04.046;%%
  %179 citations counted in INSPIRE as of 27 Aug 2017  
%\cite{Holthausen:2009uc}
%\bibitem{Holthausen:2009uc} 
  M.~Holthausen, M.~Lindner and M.~A.~Schmidt,
  %``Radiative Symmetry Breaking of the Minimal Left-Right Symmetric Model,''
  Phys.\ Rev.\ D {\bf 82}, 055002 (2010)
  doi:10.1103/PhysRevD.82.055002
  [arXiv:0911.0710 [hep-ph]];
  %%CITATION = doi:10.1103/PhysRevD.82.055002;%%
  %72 citations counted in INSPIRE as of 28 Aug 2017    
%\cite{AlexanderNunneley:2010nw}
%\bibitem{AlexanderNunneley:2010nw} 
  L.~Alexander-Nunneley and A.~Pilaftsis,
  %``The Minimal Scale Invariant Extension of the Standard Model,''
  JHEP {\bf 1009}, 021 (2010)
  doi:10.1007/JHEP09(2010)021
  [arXiv:1006.5916 [hep-ph]];
  %%CITATION = doi:10.1007/JHEP09(2010)021;%%
  %89 citations counted in INSPIRE as of 27 Aug 2017
%\cite{Hur:2011sv}
%\bibitem{Hur:2011sv} 
  T.~Hur and P.~Ko,
  %``Scale invariant extension of the standard model with strongly interacting hidden sector,''
  Phys.\ Rev.\ Lett.\  {\bf 106}, 141802 (2011)
  doi:10.1103/PhysRevLett.106.141802
  [arXiv:1103.2571 [hep-ph]];
  %%CITATION = doi:10.1103/PhysRevLett.106.141802;%%
  %101 citations counted in INSPIRE as of 27 Aug 2017  
  %\cite{Ishiwata:2011aa}
%\bibitem{Ishiwata:2011aa}
  %K.~Ishiwata,
  %``Dark Matter in Classically Scale-Invariant Two Singlets Standard Model,''
  %Phys.\ Lett.\ B {\bf 710} (2012) 134
  %doi:10.1016/j.physletb.2012.02.048
  %[arXiv:1112.2696 [hep-ph]];
  %%CITATION = doi:10.1016/j.physletb.2012.02.048;%%
  %40 citations counted in INSPIRE as of 27 Aug 2017
  %\cite{Lee:2012jn}
%\bibitem{Lee:2012jn} 
  %J.~S.~Lee and A.~Pilaftsis,
  %``Radiative Corrections to Scalar Masses and Mixing in a Scale Invariant Two Higgs Doublet Model,''
  %Phys.\ Rev.\ D {\bf 86}, 035004 (2012)
  %doi:10.1103/PhysRevD.86.035004
 % [arXiv:1201.4891 [hep-ph]];
  %%CITATION = doi:10.1103/PhysRevD.86.035004;%%
  %32 citations counted in INSPIRE as of 27 Aug 2017
%\cite{Englert:2013gz}
%\bibitem{Englert:2013gz} 
  C.~Englert, J.~Jaeckel, V.~V.~Khoze and M.~Spannowsky,
  %``Emergence of the Electroweak Scale through the Higgs Portal,''
  JHEP {\bf 1304}, 060 (2013)
  doi:10.1007/JHEP04(2013)060
  [arXiv:1301.4224 [hep-ph]];
  %%CITATION = doi:10.1007/JHEP04(2013)060;%%
  %128 citations counted in INSPIRE as of 27 Aug 2017
  %\cite{Heikinheimo:2013fta}
%\bibitem{Heikinheimo:2013fta} 
  M.~Heikinheimo, A.~Racioppi, M.~Raidal, C.~Spethmann and K.~Tuominen,
  %``Physical Naturalness and Dynamical Breaking of Classical Scale Invariance,''
  Mod.\ Phys.\ Lett.\ A {\bf 29}, 1450077 (2014)
  doi:10.1142/S0217732314500771
  [arXiv:1304.7006 [hep-ph]];
  %%CITATION = doi:10.1142/S0217732314500771;%%
  %125 citations counted in INSPIRE as of 27 Aug 2017  
    %\cite{Khoze:2013oga}
%\bibitem{Khoze:2013oga} 
  %V.~V.~Khoze and G.~Ro,
  %``Leptogenesis and Neutrino Oscillations in the Classically Conformal Standard Model with the Higgs Portal,''
  %JHEP {\bf 1310}, 075 (2013)
  %doi:10.1007/JHEP10(2013)075
  %[arXiv:1307.3764 [hep-ph]];
  %%CITATION = doi:10.1007/JHEP10(2013)075;%%
  %66 citations counted in INSPIRE as of 27 Aug 2017  
  %\cite{Carone:2013wla}
%\bibitem{Carone:2013wla} 
  C.~D.~Carone and R.~Ramos,
  %``Classical scale-invariance, the electroweak scale and vector dark matter,''
  Phys.\ Rev.\ D {\bf 88}, 055020 (2013)
  doi:10.1103/PhysRevD.88.055020
  [arXiv:1307.8428 [hep-ph]];
  %%CITATION = doi:10.1103/PhysRevD.88.055020;%%
  %89 citations counted in INSPIRE as of 27 Aug 2017
  %\cite{Farzinnia:2013pga}
%\bibitem{Farzinnia:2013pga} 
  A.~Farzinnia, H.~J.~He and J.~Ren,
  %``Natural Electroweak Symmetry Breaking from Scale Invariant Higgs Mechanism,''
  Phys.\ Lett.\ B {\bf 727}, 141 (2013)
  doi:10.1016/j.physletb.2013.09.060
  [arXiv:1308.0295 [hep-ph]];
  %%CITATION = doi:10.1016/j.physletb.2013.09.060;%%
  %95 citations counted in INSPIRE as of 27 Aug 2017
  %\cite{Dermisek:2013pta}
%\bibitem{Dermisek:2013pta} 
  D.~Chway, T.~H.~Jung, H.~D.~Kim and R.~Dermisek,
  %``Radiative Electroweak Symmetry Breaking Model Perturbative All the Way to the Planck Scale,''
  Phys.\ Rev.\ Lett.\  {\bf 113}, no. 5, 051801 (2014)
  doi:10.1103/PhysRevLett.113.051801
  [arXiv:1308.0891 [hep-ph]];
  %%CITATION = doi:10.1103/PhysRevLett.113.051801;%%
  %36 citations counted in INSPIRE as of 27 Aug 2017 
  %\cite{Antipin:2013exa}
%\bibitem{Antipin:2013exa} 
  O.~Antipin, M.~Mojaza and F.~Sannino,
  %``Conformal Extensions of the Standard Model with Veltman Conditions,''
  Phys.\ Rev.\ D {\bf 89}, no. 8, 085015 (2014)
  doi:10.1103/PhysRevD.89.085015
  [arXiv:1310.0957 [hep-ph]];
  %%CITATION = doi:10.1103/PhysRevD.89.085015;%%
  %44 citations counted in INSPIRE as of 27 Aug 2017 
  %\cite{Salvio:2014soa}
%\bibitem{Salvio:2014soa} 
  A.~Salvio and A.~Strumia,
  %``Agravity,''
  JHEP {\bf 1406}, 080 (2014)
  doi:10.1007/JHEP06(2014)080
  [arXiv:1403.4226 [hep-ph]];
  %%CITATION = doi:10.1007/JHEP06(2014)080;%%
  %103 citations counted in INSPIRE as of 27 Aug 2017
   %\cite{Allison:2014zya}
%\bibitem{Allison:2014zya} 
  K.~Allison, C.~T.~Hill and G.~G.~Ross,
  %``Ultra-weak sector, Higgs boson mass, and the dilaton,''
  Phys.\ Lett.\ B {\bf 738}, 191 (2014)
  doi:10.1016/j.physletb.2014.09.041
  [arXiv:1404.6268 [hep-ph]];
  %%CITATION = doi:10.1016/j.physletb.2014.09.041;%%
  %33 citations counted in INSPIRE as of 27 Aug 2017
%\cite{Lindner:2014oea}
%\bibitem{Lindner:2014oea} 
  %M.~Lindner, S.~Schmidt and J.~Smirnov,
  %``Neutrino Masses and Conformal Electro-Weak Symmetry Breaking,''
  %JHEP {\bf 1410}, 177 (2014)
  %doi:10.1007/JHEP10(2014)177
  %[arXiv:1405.6204 [hep-ph]];
  %%CITATION = doi:10.1007/JHEP10(2014)177;%%
  %59 citations counted in INSPIRE as of 27 Aug 2017
 %\cite{Altmannshofer:2014vra}
%\bibitem{Altmannshofer:2014vra} 
  W.~Altmannshofer, W.~A.~Bardeen, M.~Bauer, M.~Carena and J.~D.~Lykken,
  %``Light Dark Matter, Naturalness, and the Radiative Origin of the Electroweak Scale,''
  JHEP {\bf 1501}, 032 (2015)
  doi:10.1007/JHEP01(2015)032
  [arXiv:1408.3429 [hep-ph]];
  %%CITATION = doi:10.1007/JHEP01(2015)032;%%
  %45 citations counted in INSPIRE as of 27 Aug 2017 
  %\cite{Wang:2015cda}
%\bibitem{Wang:2015cda} 
  Z.~W.~Wang, T.~G.~Steele, T.~Hanif and R.~B.~Mann,
  %``Conformal Complex Singlet Extension of the Standard Model: Scenario for Dark Matter and a Second Higgs Boson,''
  JHEP {\bf 1608}, 065 (2016)
  doi:10.1007/JHEP08(2016)065
  [arXiv:1510.04321 [hep-ph]];
  %%CITATION = doi:10.1007/JHEP08(2016)065;%%
  %11 citations counted in INSPIRE as of 27 Aug 2017
  %\cite{Ghorbani:2015xvz}
%\bibitem{Ghorbani:2015xvz} 
  %K.~Ghorbani and H.~Ghorbani,
  %``Scalar Dark Matter in Scale Invariant Standard Model,''
  %JHEP {\bf 1604}, 024 (2016)
  %doi:10.1007/JHEP04(2016)024
  %[arXiv:1511.08432 [hep-ph]];
  %%CITATION = doi:10.1007/JHEP04(2016)024;%%
  %11 citations counted in INSPIRE as of 27 Aug 2017 
%\cite{Haba:2015qbz}
%\bibitem{Haba:2015qbz} 
  N.~Haba, H.~Ishida, N.~Kitazawa and Y.~Yamaguchi,
  %``A new dynamics of electroweak symmetry breaking with classically scale invariance,''
  Phys.\ Lett.\ B {\bf 755}, 439 (2016)
  doi:10.1016/j.physletb.2016.02.052
  [arXiv:1512.05061 [hep-ph]];
  %%CITATION = doi:10.1016/j.physletb.2016.02.052;%%
  %17 citations counted in INSPIRE as of 27 Aug 2017
  %\cite{Helmboldt:2016mpi}
%\bibitem{Helmboldt:2016mpi} 
  A.~J.~Helmboldt, P.~Humbert, M.~Lindner and J.~Smirnov,
  %``Minimal conformal extensions of the Higgs sector,''
  JHEP {\bf 1707}, 113 (2017)
  doi:10.1007/JHEP07(2017)113
  [arXiv:1603.03603 [hep-ph]];
  %%CITATION = doi:10.1007/JHEP07(2017)113;%%
  %11 citations counted in INSPIRE as of 27 Aug 2017
  %\cite{Ahriche:2016ixu}
%\bibitem{Ahriche:2016ixu} 
  A.~Ahriche, A.~Manning, K.~L.~McDonald and S.~Nasri,
  %``Scale-Invariant Models with One-Loop Neutrino Mass and Dark Matter Candidates,''
  Phys.\ Rev.\ D {\bf 94}, no. 5, 053005 (2016)
  doi:10.1103/PhysRevD.94.053005
  [arXiv:1604.05995 [hep-ph]];
  %%CITATION = doi:10.1103/PhysRevD.94.053005;%%
  %20 citations counted in INSPIRE as of 27 Aug 2017
  %\cite{Khoze:2016zfi}
%\bibitem{Khoze:2016zfi} 
  %V.~V.~Khoze and A.~D.~Plascencia,
  %``Dark Matter and Leptogenesis Linked by Classical Scale Invariance,''
  %JHEP {\bf 1611}, 025 (2016)
  %doi:10.1007/JHEP11(2016)025
  %[arXiv:1605.06834 [hep-ph]];
  %%CITATION = doi:10.1007/JHEP11(2016)025;%%
  %10 citations counted in INSPIRE as of 27 Aug 2017
  %\cite{Karam:2016rsz}
%\bibitem{Karam:2016rsz} 
  A.~Karam and K.~Tamvakis,
  %``Dark Matter from a Classically Scale-Invariant $SU(3)_X$,''
  Phys.\ Rev.\ D {\bf 94}, no. 5, 055004 (2016)
  doi:10.1103/PhysRevD.94.055004
  [arXiv:1607.01001 [hep-ph]];
  %%CITATION = doi:10.1103/PhysRevD.94.055004;%%
  %18 citations counted in INSPIRE as of 27 Aug 2017 
  %\cite{Abel:2017ujy}
%\bibitem{Abel:2017ujy} 
  S.~Abel and F.~Sannino,
  %``Radiative symmetry breaking from interacting UV fixed points,''
  arXiv:1704.00700 [hep-ph].
  %%CITATION = ARXIV:1704.00700;%%
  %7 citations counted in INSPIRE as of 27 Aug 2017

%\cite{Kobakhidze:2017eml}
\bibitem{Kobakhidze:2017eml} 
  A.~Kobakhidze and S.~Liang,
  %``Standard Model with hidden scale invariance and light dilaton,''
  arXiv:1701.04927 [hep-ph].
  %%CITATION = ARXIV:1701.04927;%%
  %2 citations counted in INSPIRE as of 01 Sep 2017

%\cite{Kobakhidze:2017mcz}
\bibitem{Kobakhidze:2017mcz} 
  A.~Kobakhidze and S.~Liang,
  %``Scale Invariant Top Condensate,''
  arXiv:1707.05942 [hep-ph].
  %%CITATION = ARXIV:1707.05942;%%

%\cite{Witten:1980ez}
\bibitem{Witten:1980ez} 
  E.~Witten,
  %``Cosmological Consequences of a Light Higgs Boson,''
  Nucl.\ Phys.\ B {\bf 177}, 477 (1981).
  doi:10.1016/0550-3213(81)90182-6
  %%CITATION = doi:10.1016/0550-3213(81)90182-6;%%
  %159 citations counted in INSPIRE as of 01 Sep 2017
  
  %\cite{Buchmuller:1990ds}
\bibitem{Buchmuller:1990ds} 
  W.~Buchmuller and D.~Wyler,
  %``The Effect of dilatons on the electroweak phase transition,''
  Phys.\ Lett.\ B {\bf 249}, 281 (1990).
  doi:10.1016/0370-2693(90)91256-B
  %%CITATION = doi:10.1016/0370-2693(90)91256-B;%%
  %13 citations counted in INSPIRE as of 01 Sep 2017

%\cite{Iso:2017uuu}
\bibitem{Iso:2017uuu} 
  S.~Iso, P.~D.~Serpico and K.~Shimada,
  %``QCD-Electroweak first order phase transition in supercooled universe,''
  arXiv:1704.04955 [hep-ph].
  %%CITATION = ARXIV:1704.04955;%%
  %1 citations counted in INSPIRE as of 27 Aug 2017   
  
  %\cite{Huynh:2013aea}
\bibitem{Huynh:2013aea} 
  M.~Huynh and J.~Lazio,
  %``An Overview of the Square Kilometre Array,''
  arXiv:1311.4288 [astro-ph.IM].
  %%CITATION = ARXIV:1311.4288;%%
  %3 citations counted in INSPIRE as of 07 Sep 2017
  
   %\cite{Shaposhnikov:2008xi}
\bibitem{Shaposhnikov:2008xi} 
  M.~Shaposhnikov and D.~Zenhausern,
  %``Quantum scale invariance, cosmological constant and hierarchy problem,''
  Phys.\ Lett.\ B {\bf 671}, 162 (2009)
  doi:10.1016/j.physletb.2008.11.041
  [arXiv:0809.3406 [hep-th]];
  %%CITATION = doi:10.1016/j.physletb.2008.11.041;%%
  %124 citations counted in INSPIRE as of 04 Jan 2017
   %\cite{Ghilencea:2016dsl}
%\bibitem{Ghilencea:2016dsl} 
  D.~M.~Ghilencea, Z.~Lalak and P.~Olszewski,
  %``Standard Model with spontaneously broken quantum scale invariance,''
  arXiv:1612.09120 [hep-ph].
  %%CITATION = ARXIV:1612.09120;%%
  
  %\cite{Foot:2010et}
\bibitem{Foot:2010et} 
  R.~Foot, A.~Kobakhidze and R.~R.~Volkas,
  %``Cosmological constant in scale-invariant theories,''
  Phys.\ Rev.\ D {\bf 84}, 075010 (2011)
  doi:10.1103/PhysRevD.84.075010
  [arXiv:1012.4848 [hep-ph]];
  %%CITATION = doi:10.1103/PhysRevD.84.075010;%%
  %21 citations counted in INSPIRE as of 19 Nov 2016
  %\cite{Foot:2011et}
%\bibitem{Foot:2011et} 
  R.~Foot and A.~Kobakhidze,
  %``Electroweak Scale Invariant Models with Small Cosmological Constant,''
  Int.\ J.\ Mod.\ Phys.\ A {\bf 30}, no. 21, 1550126 (2015)
  doi:10.1142/S0217751X15501262
  [arXiv:1112.0607 [hep-ph]].
  %%CITATION = doi:10.1142/S0217751X15501262;%%
  %8 citations counted in INSPIRE as of 19 Nov 2016

%\cite{Kobakhidze:2013pya}
\bibitem{Kobakhidze:2013pya} 
  A.~Kobakhidze and A.~Spencer-Smith,
  %``Neutrino Masses and Higgs Vacuum Stability,''
  JHEP {\bf 1308}, 036 (2013)
  doi:10.1007/JHEP08(2013)036
  [arXiv:1305.7283 [hep-ph]].
  %%CITATION = doi:10.1007/JHEP08(2013)036;%%
  %22 citations counted in INSPIRE as of 03 Sep 2017
  
  %\cite{Foot:2007wn}
\bibitem{Foot:2007wn} 
  R.~Foot, A.~Kobakhidze, K.~L.~McDonald and R.~R.~Volkas,
  %``Renormalization-scale independence of the physical cosmological constant,''
  Phys.\ Lett.\ B {\bf 664}, 199 (2008)
  doi:10.1016/j.physletb.2008.05.029
  [arXiv:0712.3040 [hep-th]].
  %%CITATION = doi:10.1016/j.physletb.2008.05.029;%%
  %12 citations counted in INSPIRE as of 03 Sep 2017
  
  %\cite{Quiros:1999jp}
\bibitem{Quiros:1999jp} 
  M.~Quiros,
  %``Finite temperature field theory and phase transitions,''
  hep-ph/9901312.
  %%CITATION = HEP-PH/9901312;%%
  %199 citations counted in INSPIRE as of 04 Sep 2017
  
  %\cite{Gasser:1986vb}
\bibitem{Gasser:1986vb} 
  J.~Gasser and H.~Leutwyler,
  %``Light Quarks at Low Temperatures,''
  Phys.\ Lett.\ B {\bf 184}, 83 (1987).
  doi:10.1016/0370-2693(87)90492-8
  %%CITATION = doi:10.1016/0370-2693(87)90492-8;%%
  %615 citations counted in INSPIRE as of 07 Sep 2017
  
  %\cite{Pisarski:1983ms}
\bibitem{Pisarski:1983ms} 
  R.~D.~Pisarski and F.~Wilczek,
  %``Remarks on the Chiral Phase Transition in Chromodynamics,''
  Phys.\ Rev.\ D {\bf 29}, 338 (1984).
  doi:10.1103/PhysRevD.29.338
  %%CITATION = doi:10.1103/PhysRevD.29.338;%%
  %1170 citations counted in INSPIRE as of 07 Sep 2017
  
  %\cite{Gausterer:1988fv}
\bibitem{Gausterer:1988fv} 
  H.~Gausterer and S.~Sanielevici,
  %``Can the Chiral Transition in {QCD} Be Described by a Linear $\sigma$ Model in Three-dimensions?,''
  Phys.\ Lett.\ B {\bf 209}, 533 (1988).
  doi:10.1016/0370-2693(88)91188-4
  %%CITATION = doi:10.1016/0370-2693(88)91188-4;%%
  %19 citations counted in INSPIRE as of 07 Sep 2017
  
  %\cite{Caprini:2010xv}
\bibitem{Caprini:2010xv} 
  C.~Caprini, R.~Durrer and X.~Siemens,
  %``Detection of gravitational waves from the QCD phase transition with pulsar timing arrays,''
  Phys.\ Rev.\ D {\bf 82}, 063511 (2010)
  doi:10.1103/PhysRevD.82.063511
  [arXiv:1007.1218 [astro-ph.CO]].
  %%CITATION = doi:10.1103/PhysRevD.82.063511;%%
  %37 citations counted in INSPIRE as of 07 Sep 2017
  
  %\cite{Kobakhidze:2017mru}
\bibitem{Kobakhidze:2017mru} 
  A.~Kobakhidze, C.~Lagger, A.~Manning and J.~Yue,
  %``Gravitational waves from a supercooled electroweak phase transition and their detection with pulsar timing arrays,''
  Eur.\ Phys.\ J.\ C {\bf 77}, no. 8, 570 (2017)
  doi:10.1140/epjc/s10052-017-5132-y
  [arXiv:1703.06552 [hep-ph]].
  %%CITATION = doi:10.1140/epjc/s10052-017-5132-y;%%
  %5 citations counted in INSPIRE as of 07 Sep 2017
  
%\cite{Crawford:1982yz}
\bibitem{Crawford:1982yz} 
  M.~Crawford and D.~N.~Schramm,
  %``Spontaneous Generation of Density Perturbations in the Early Universe,''
  Nature {\bf 298}, 538 (1982).
  doi:10.1038/298538a0;
  %%CITATION = doi:10.1038/298538a0;%%
  %135 citations counted in INSPIRE as of 07 Sep 2017 
   %\cite{Hall:1989hr}
%\bibitem{Hall:1989hr} 
  L.~J.~Hall and S.~Hsu,
  %``Cosmological Production of Black Holes,''
  Phys.\ Rev.\ Lett.\  {\bf 64}, 2848 (1990).
  doi:10.1103/PhysRevLett.64.2848;
  %%CITATION = doi:10.1103/PhysRevLett.64.2848;%%
  %21 citations counted in INSPIRE as of 07 Sep 2017
  %\cite{Jedamzik:1996mr}
%\bibitem{Jedamzik:1996mr} 
  K.~Jedamzik,
  %``Primordial black hole formation during the QCD epoch,''
  Phys.\ Rev.\ D {\bf 55}, 5871 (1997)
  doi:10.1103/PhysRevD.55.5871
  [astro-ph/9605152].
  %%CITATION = doi:10.1103/PhysRevD.55.5871;%%
  %122 citations counted in INSPIRE as of 07 Sep 2017

%\cite{Servant:2014bla}
\bibitem{Servant:2014bla} 
  G.~Servant,
  %``Baryogenesis from Strong $CP$ Violation and the QCD Axion,''
  Phys.\ Rev.\ Lett.\  {\bf 113}, no. 17, 171803 (2014)
  doi:10.1103/PhysRevLett.113.171803
  [arXiv:1407.0030 [hep-ph]].
  %%CITATION = doi:10.1103/PhysRevLett.113.171803;%%
  %18 citations counted in INSPIRE as of 07 Sep 2017
  
 % \cite{new}
\bibitem{new} 
  S.~Arunasalam, A.~Kobakhidze, C.~Lagger and A.~Zhou, ``Cosmological implications of hidden scale invariance,'' work in progress.


\end{thebibliography}
\end{document}